\documentclass{aa}  
\pdfoutput=1
\usepackage{graphicx}
\usepackage{txfonts}
\usepackage[
    colorlinks=true,       
    linkcolor=blue,        
    citecolor=blue,        
    filecolor=magenta,     
    urlcolor=blue,         
    pdfborder={0 0 0},     
    bookmarks=true,        
    bookmarksopen=true,    
    bookmarksnumbered=true,
    hyperfootnotes=true,   
    breaklinks=true        
]{hyperref}
\usepackage[dvipsnames]{xcolor}
\usepackage{bm}

\newcommand{\eqRef}[1]{equation \ref{#1}}

\newcommand{\figRef}[1]{figure \ref{#1}}
\newcommand{\FIGRef}[1]{Figure \ref{#1}}

\begin{document}

   \title{Toward Realistic Solar Flare Models}
   \subtitle{An explicit Particle-In-Cell solver in the DISPATCH framework}

   \author{M. Haahr
          \inst{1,2}
          \and
          B. V. Gudiksen\inst{1,2}
          \and
          Å. Nordlund \inst{1,3}
          }

   \institute{ Rosseland Centre for Solar Physics, University of Oslo, P.O. Box 1029 Blindern, NO-0315 Oslo, Norway\\
    \email{michhaa@uio.no}
         \and Institute of Theoretical Astrophysics, University of Oslo, P.O. Box 1029 Blindern, NO-0315 Oslo, Norway \\
         \and  Centre for Star and Planet Formation, Niels Bohr Institute, University of Copenhagen, Øster Voldgade 5-7, 1350 Copenhagen, Denmark\\
             }

   \date{\today}
 
  \abstract
   {Simulating solar flares, which involve large-scale dynamics and small-scale magnetic reconnection, poses significant computational challenges.}
   {This study aims to develop an explicit Particle-In-Cell (PIC) solver within the DISPATCH framework to model the small-scale kinetic processes in solar corona setting. This study in the first in a series with the ultimate goal to develop a hybrid PIC-MHD solver, to simulate solar flares.}
   {The PIC solver, inspired by the PhotonPlasma code, solves the Vlasov-Maxwell equations in a collisionless regime using explicit time-staggering and spatial-staggering techniques. Validation included unit tests, plasma frequency recovery, two-stream instability, and current sheet dynamics.}
   {Validation tests confirmed the solver's accuracy and robustness in modeling plasma dynamics and electromagnetic fields.}
   {The integration of the explicit PIC solver into the DISPATCH framework is the first step towards bridging the gap between large and small scale dynamics, providing a robust platform for future solar physics research.}


   \maketitle
\section{Introduction}
Simulating solar flares, encompassing both the macroscopic evolution and the microscopic details of current sheets that form during magnetic reconnection events, remains an unsolved computational challenge. According to recent work 
\citep{guoMagneticReconnectionAssociated2023a}, no computational tool currently bridges the gap between the large-scale dynamics of solar flares and the intricate processes within magnetic reconnection regions.

Solar flares are generally accepted to result from the release of magnetic energy in magnetic reconnection regions \citep{zharkovaRecentAdvancesUnderstanding2011}. Within these regions, some particles are accelerated to non-thermal, relativistic speeds. These high-energy particles exchange energy with the surrounding plasma. The non-local effects of these particles influence plasma behavior on large scales, which in turn affects the small scale dynamics. This interplay between large and small scales is critical for understanding the initiation and development of solar flares. 

Kinetic approaches, particularly the Particle-In-Cell (PIC) method  \citep{liuFirstPrinciplesTheoryRate2022a, klionParticleincellSimulationsRelativistic2023a, selviEffectiveResistivityRelativistic2023}, have been extensively applied to study reconnection at microscales. Despite their utility, these studies often resort to simplified, idealized configurations, such as the 2D Harris sheet model \citep{harrisPlasmaSheathSeparating1962}, due to computational constraints. Given that real-world solar flares can extend over vast distances, employing PIC methods exclusively becomes impractical. Although there have been efforts to combine Magnetohydrodynamics (MHD) and PIC techniques \citep{baumannKINETICMODELINGPARTICLE2013a,daldorffTwowayCouplingGlobal2014a}, challenges persist in accurately capturing the multiscale nature of solar flares. A hybrid modeling approach that can seamlessly integrate different temporal and spatial scales is thus paramount for realistic simulations.

Handling the wide range of temporal and spatial scales in solar flare simulations, while maintaining computational performance, is a significant challenge. Developing a sophisticated multi-scale, multiphysics solver from scratch would be daunting. Fortunately, the DISPATCH framework \citep{nordlundDISPATCHNumericalSimulation2018b} provides a robust foundation for handling the scales required to simulate solar flares. DISPATCH organizes the simulation domain into "patches," each updated semi-autonomously. This localized data management enables each patch to function as a unique type of solver within a simulation, facilitating simulations with multiphysics solvers. DISPATCH has already proven capable of modeling complex, multiscale environments, such as the entire solar convection zone \citep{popovasGlobalMHDSimulations2022a}.

This paper marks the initial phase in developing a hybrid solver within the DISPATCH framework by integrating a PIC solver inspired by the PhotonPlasma code \citep{haugbollePhotonplasmaModernHighorder2013a}. We detail the process of embedding the PIC solver into DISPATCH, noting the essential modifications made from the original PhotonPlasma code to ensure compatibility and effectiveness within the DISPATCH architecture. The PIC solver is tailored for future integration with MHD simulations. Through this integration, we aim to bridge the gap between large-scale flare dynamics and the detailed kinetic processes occurring within reconnection zones, thereby contributing to a more comprehensive understanding of solar flare mechanisms.

\section{Methods}

\subsection{Governing Equations}
\label{sec:equations}

The PIC solver implemented in this study solves the Vlasov-Maxwell system of equations. 

For the sake of reproducibility and clarity, we outline only the most important equations here. For an exhaustive description, readers are referred to the original description of the PhotonPlasma code \citep{haugbollePhotonplasmaModernHighorder2013a}. In our notation, we use bold letters to indicate vectors. The Vlasov-Maxwell system is represented as follows:
\begin{equation}
    \frac{\partial \mathbf{f}^s}{\partial t } + \mathbf{u} \cdot \frac{\partial \mathbf{f}^s}{\partial \mathbf{x}} + \frac{q^s}{m^s} (\mathbf{E} + k_F \mathbf{u} \times \mathbf{B}) \cdot \frac{\partial \mathbf{f}^s}{\partial (\mathbf{u} \gamma)} = C
    \label{eq:Vlasov}
\end{equation}
Here, $\mathbf{f}$ is the distribution function, which is a function of time, velocity ($\mathbf{u}$), and space ($\mathbf{x}$). $s$ denotes particle species (e.g., electrons, protons), $\gamma = {1}/{\sqrt{1 - \left({u}/{c}\right)^2}}$ the relativistic Lorentz factor and $C$ the collision operator. Finally, $\mathbf{E}$ and $\mathbf{B}$ are the electric and magnetic fields respectively. For this study, we focus on the collisionless regime, setting $C=0$. 

The electric and magnetic fields' evolution follow the Maxwell equations:
\begin{equation}
    \label{eq:Ampere_law}
    \frac{\partial \mathbf{E}}{\partial t} = \frac{k_E}{k_B} \mathbf{\nabla} \times \mathbf{B} - 4\pi k_E \mathbf{J} 
\end{equation}
\begin{equation}
    \label{eq:Faraday_law}
    \frac{\partial \mathbf{B}}{\partial t} = - \frac{1}{k_F}  \mathbf{\nabla} \times \mathbf{E},
\end{equation}
\begin{equation}
    \label{eq:div_B}
    \mathbf{\nabla} \cdot \mathbf{B} = 0,
\end{equation}
\begin{equation}
    \label{eq:div_E}
    \mathbf{\nabla} \cdot \mathbf{E} = k_E 4 \pi \rho_c,
\end{equation}
where $\mathbf{J}$ and $\rho_c$ represent the current density and charge density, respectively. 

The constants $k_F$ $k_E$, and $k_B$ ensure the equations' consistency across different unit systems, as detailed by \citet{carronBabelUnitsEvolution2015a}. We show the values corresponding to SI, CGS, and HL units in table \ref{tab:Lorentz-Maxwell}. The use of this unit-agnostic formulation is commented on in section \ref{sec:unit_agnostic}

The PIC solver approximates the Vlasov equation (\eqRef{eq:Vlasov}) by sampling the phase space with "macro particles", each representing a collective of real particles, as discussed in works such as \citet{birdsallPlasmaPhysicsComputer1991} and \citet{lapentaKineticApproachMicroscopicmacroscopic2006}. Macro particles are assigned a weight that represents the number of real particles they sample. The macro particles are then treated as individual particles, with their velocities influenced by the Lorentz force:

\begin{equation}
    \frac{d (\mathbf{u}_p \gamma)}{dt} = \frac{q_s}{m_s} (\mathbf{E} + k_F \mathbf{u}_p \times \mathbf{B}),
    \label{eq:Lorentz_foce}
\end{equation}
where their positions evolve as per:
\begin{equation}
    \frac{d \mathbf{x_p}}{dt} = \mathbf{u}_p.
    \label{eq:particle_position}
\end{equation}
Here $\mathbf{u}_p$ and $\mathbf{x}_p$ is the velocity and position of each individual macro particle. $q_s$ and $m_s$ is the charge and mass of the particle species.

Interpolating electric and magnetic fields to macro particle positions require assigning specific shapes to these particles, affecting field interpolation. Following PhotonPlasma's methodology, we have implemented various shape functions, including Nearest-Grid-Point (NGP), Cloud-in-Cell (CIC), Triangular-Shaped-Cloud (TSC), and Piecewise-Cubic-Spline (PCS), noting that NGP can introduce instabilities \citep{birdsallPlasmaPhysicsComputer1991,smetsNewMethodDispatch2021}. We use 'CIC' interpolation and 2nd order derivatives as the default in our solver. 

Determining properties such as charge density from particle distributions uses similar shape function, and we refer to this as 'particle deposition' in this article. The interpolation and deposition shape functions can be defined independently as simulation input parameters.

\begin{table}[h]
\centering
\renewcommand{\arraystretch}{1.5} 
\begin{tabular}{|c|c|c|c|}
\hline
 & SI & Gaussian (CGS) & Heaviside-Lorentz (HL)\\
\hline
\hline
$k_E$ & $\frac{1}{4 \pi \epsilon_0}$ & $1$ & $\frac{1}{4\pi}$ \\
$k_B$ & $\frac{\mu_0}{4 \pi}$ & $\frac{1}{c}$ & $\frac{1}{4 \pi c}$\\
$k_F$ & $1$ & $\frac{1}{c}$ & $\frac{1}{c}$ \\
\hline
\end{tabular}
\caption{Constants used in the Maxwell-Lorentz equations for system-agnostic formulation.}
\label{tab:Lorentz-Maxwell}
\end{table}

\subsection{Transparent Scaling}
\subsubsection{Scaling}
Scaling of simulation parameters plays a critical role in ensuring the stability and optimizing the performance of simulations. Although frequently mentioned in scientific literature, the detailed methodologies for scaling are often not fully elaborated, posing challenges to reproducibility. In response to this, we aim to clearly outline the scaling techniques employed in our study.

Within the DISPATCH framework, we have incorporated a scaling module to handle scaling for SI, CGS, and HL unit systems. This module relies on three primary parameters for scaling: \textbf{mass density}, \textbf{length}, and \textbf{time}. The scaling of all subsequent units stems from these foundational parameters. 

Often, variables of interest such as the proton mass, the elementary charge, and the speed of light, are set to 1 in code units. Along with this, pressure, density, and field strength are usually given in code units as well, making it difficult to recover the 'real' physical units. For ease of physical interpretation and to ensure that input parameters remain intuitive, we have based all scaling on real physical units. As such, the code is designed to accept input parameters in familiar terms, including common length scales, magnetic field strengths, number densities, and temperatures.

\subsubsection{Fudging}

A common practice in PIC simulations is the "fudging" of physical constants to adjust time and length scales appropriately. This process, although critical for the fine-tuning of simulations, is frequently glossed over in the literature, leaving a gap in understanding regarding the specific adjustments made. Our goal is to demystify this aspect by ensuring complete transparency in our scaling process. 

In our simulations, we permit the modification of three key constants: the \textbf{speed of light}, \textbf{electron mass}, and \textbf{elementary charge}. The modification of the speed of light is facilitated through changes to the vacuum permeability, $\mu_0$, or  vacuum permittivity, $\epsilon_0$. Specifically, in CGS and HL units where $\mu_0$ is dimensionless, alterations to the speed of light are by adjusting $\epsilon_0$. 

\subsubsection{Unit-Agnostic Equations}
\label{sec:unit_agnostic}

The Vlasov-Maxwell system of equations varies depending on the unit system used. Often, the unit system is not explicitly stated and the equations written in a 'natural units' where factors such as $4\pi$ are omitted. Gaussian and Heaviside-Lorentz (HL) are examples of 'natural units' and of often both referred to as CGS. Although units may appear identical in Gaussian and HL units their actual values differ. For instance, a magnetic field strength of $50G$ in Gaussian units corresponds to approximately $14.1G$ in HL units, differing by a factor $\sqrt{4\pi}$. While an experienced reader may deduce the underlying system, this can be problematic for inexperienced readers. This can make it challenging to reproduce results if the exact unit system is not specified.

'Fudging' the physical constants also alters the equations. When ambiguous fudging is combined with an ambiguous unit system, it presents a complex puzzle for the reader to solve.

To assist the reader, we have written all equations in a unit-agnostic form. Constants are included in the equations, and these constants vary depending on the unit system and the 'fudging' used, as shown in Table \ref{tab:Lorentz-Maxwell}. This approach allows us to use the same equations in both 'real' units and code units, simplifying implementation and improving the clarity of our code.

\subsection{Numerical Approach}

\subsubsection{DISPATCH}

A fundamental understanding of the DISPATCH framework is needed to understand some of our choices made when implementing the PIC code into DISPATCH. While a detailed description  of DISPATCH can be found in \citet{nordlundDISPATCHNumericalSimulation2018b}, we here summarize the key aspects relevant to our PIC solver integration.

DISPATCH organizes the simulation domain into distinct sections referred to as "patches". These patches are updated semi-autonomously, interacting with adjacent patches primarily to update ghost zones. This architecture offers substantial benefits, specifically the ability for each patch to proceed with its own timestep. This feature significantly boosts the potential for simulation efficiency and enables the simultaneous resolution of diverse timescales within a single experiment.

The framework's design, focusing on localized data management, inherently limits the applicability of global convergence schemes like implicit methods. Furthermore, DISPATCH imposes specific restrictions on the communication between patches, designed to manage the complexity of data interactions efficiently. 


Adapting to DISPATCH's focus on localized computations, our implementation utilizes an explicit PIC solver, staggered in both time and space. This choice is compatible with DISPATCH's decentralized data handling paradigm, ensuring computational efficiency and precision while navigating the framework's structural nuances.

\subsubsection{Discretization}

\begin{figure}[]
    \centering
    \includegraphics[width=0.45\textwidth]{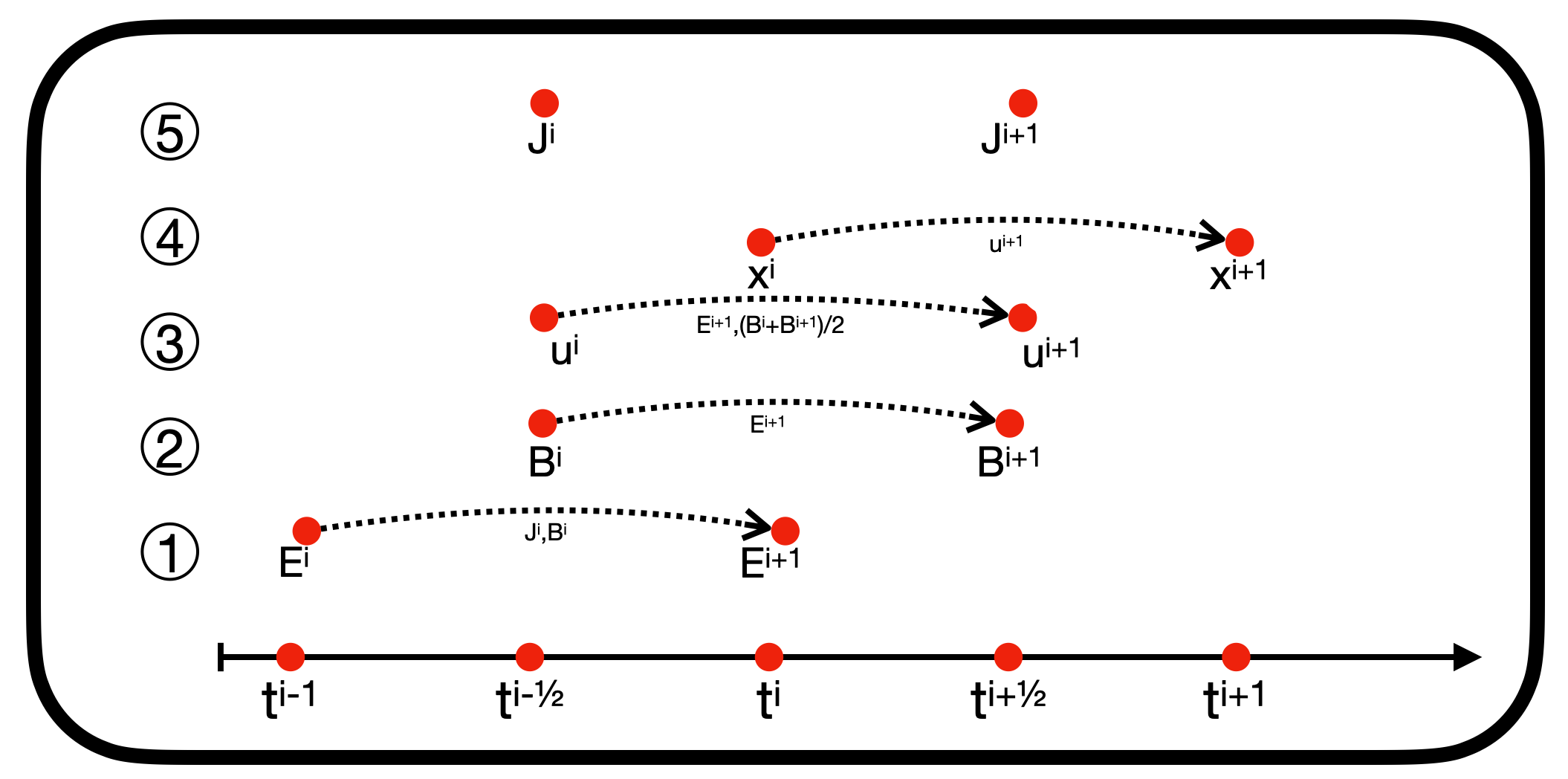}
    \caption{Time staggering of the explicit PIC solver in DISPATCH, illustrating the integration order. Electric field are staggered by a full timestep,  $\Delta t$ in the past. Magnetic field, particle velocities, and current density are staggered backwards by $0.5\Delta t$. Particle position is centered in time.}
    \label{fig:time_stagger}
\end{figure}

\begin{figure}
    \centering
    \includegraphics[width=0.45\textwidth]{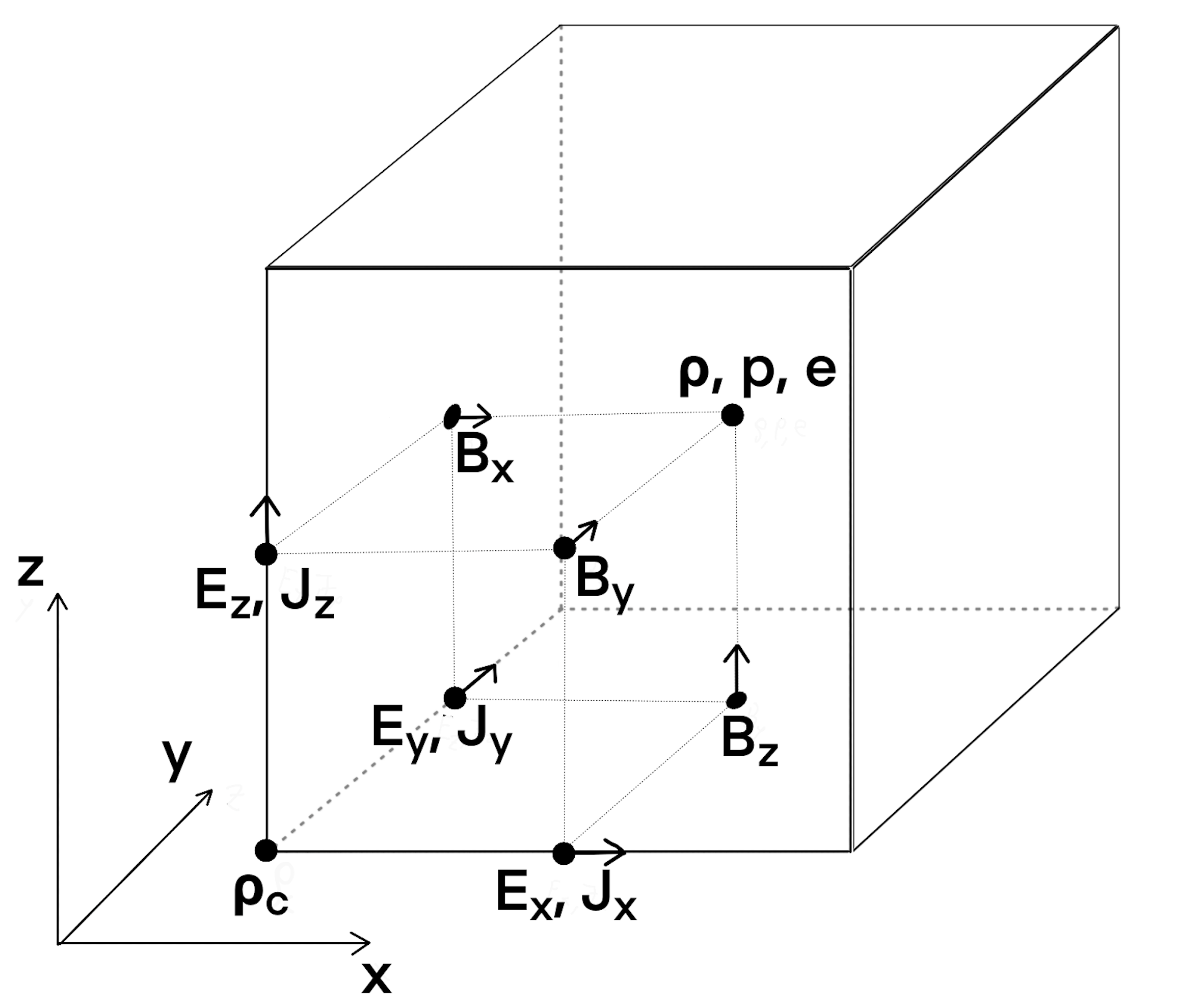}
    \caption{Spatial staggering of mesh variables on a Yee lattice. Magnetic fields are located on the cell faces, electric fields and current densities on the edges, with mass density, momentum, and energy centered within each cell. Charge density is uniquely staggered to the lower corner of each cell.}
    \label{fig:space_stagger}
\end{figure}

For the numerical stability and reliability of our explicit computational scheme, we incorporate a time-staggering method akin to the leap-frog integration technique, as visually represented in \figRef{fig:time_stagger}. The leap-frog scheme (\cite{feynmanFeynmanLecturesPhysics1965} section 9.6, \cite{birdsallPlasmaPhysicsComputer1991} p. 56) is renowned for its efficacy in maintaining stability during oscillatory motion— a common occurrence within PIC simulations. The electric field is staggered by a full timestep,  $\Delta t$ in the past. Magnetic field, particle velocities, and current density are staggered backwards by $0.5\Delta t$. Particle position is centered in time.

Spatially, we stagger mesh variables in accordance with a Yee lattice setup (\cite{kaneyeeNumericalSolutionInitial1966}), to align with DISPATCH standards. This spatial organization involves positioning magnetic fields on the cell faces, electric fields and current densities are staggered along cell edges. Consistent with DISPATCH conventions, mass density, momentum, and energy metrics are centered within each cell. Uniquely, charge density is allocated to the lower corner of each cell, as shown in \figRef{fig:space_stagger}. 

This arrangement of variables in time and space minimizes the interpolation needed, thereby improving the simulation's computational efficiency and accuracy.

In our simulation, both electric and magnetic fields are interpolated from the mesh to each particle's position. Interpolating fields to particle positions involves considerations of both time and space. As outlined in \figRef{fig:time_stagger}, the magnetic field requires time interpolation between the timestep $i$, and timestep $i+1$. However, with varying timesteps this introduces the potential for this time interpolation to become desynchronized. Within the DISPATCH framework, each patch dynamically sets its timestep based on the local maximum velocity to comply with the Courant condition \citep{courantUberPartiellenDifferenzengleichungen1928}. This approach is applied to all patches in the simulation, including the PIC solver patches. Due to the PIC solver's focus on electromagnetic phenomena and the need to accurately simulate the propagation of electromagnetic waves at the speed of light, the "maximum velocity" in PIC patches defaults to the speed of light. This condition results in a uniform timestep across these patches. Consequently, time interpolation for the magnetic field is simplified to averaging $B^i$ and $B^{i+1}$, as outlined in \figRef{fig:time_stagger}.

Spatially, after addressing time interpolation, fields are then interpolated to the exact spatial locations of particles. This spatial interpolation relies on the shape functions introduced in Section \ref{sec:equations}.


Particle deposition into current density, mass density, and other mesh variables happens after the particles have been updated, as described in section \ref{sec:charge_density}. This process ensures that changes in particle distribution are accurately reflected in the mesh. Consequently, electromagnetic field calculations can incorporate the most current state of the plasma. We optionally allow the shape function for deposition to differ from the one used for interpolation, but as default the shape functions are identical.

\subsubsection{Solving Maxwell's Equations}
The update cycle for each patch begins with updating electric and magnetic fields. The integration of Maxwell's equations in our solver uses an explicit method, sequentially updating the electric and magnetic fields.
The updating process initiates by updating $E^i$ to $E^{i+1}$, (\figRef{fig:time_stagger}), using \eqRef{eq:Ampere_law}. This calculation uses the magnetic field, $B^i$,
alongside the current density, $J^{i}$. The magnetic field is then updated in line with \eqRef{eq:Faraday_law}, using the just-updated electric field.


Accumulation of errors over time can require the use of 'cleaners' for both the electric and magnetic fields. Within the DISPATCH framework, divergence cleaners for the magnetic field are employed to uphold \eqRef{eq:div_B} in the guard zones, where time-interpolation in general does not conserve div(B) exactly. A similar cleaning strategy is applied to the electric field to ensure it meets the requirements of \eqRef{eq:div_E}. This cleaning procedure begins by identifying the divergence error, \(\epsilon\), through the calculation of the electric field's divergence, from which the charge density is then subtracted:
\begin{equation}
    \epsilon = \mathbf{\nabla} \cdot \mathbf{E} - k_E 4 \pi \rho_c.
\end{equation}
A subsequent Poisson filtering step addresses only the errors local to each patch to compute 
\(\Phi_{\epsilon}\) from 
\begin{equation}\label{eq:PhiE}
  \Delta\Phi_{\epsilon} = \epsilon , 
\end{equation}
after which the electric field is updated to a 'cleaned' state:
\begin{equation}\label{eq:EPhi}
    \mathbf{E}_{clean} = \mathbf{E} - \mathbf{\nabla} \Phi_{\epsilon}.
\end{equation}

The implementation allows users to decide the frequency of the electric field cleaning step and whether to use it at all. 


\subsubsection{Particle Movement}
Next in the update cycle, particle velocities are updated, affecting both particles in the "inner" domain and those within the ghost zones. As highlighted earlier, the electric and magnetic fields are interpolated to the particles' positions before their velocities are updated.

The update of velocities uses the Lorentz force equation (\eqRef{eq:Lorentz_foce}). \citet{ripperdaComprehensiveComparisonRelativistic2018} compares several different explicit approaches to implement this numerically. We have chosen to use the Vay particle pusher algorithm \citep{vaySimulationBeamsPlasmas2008a} due to its ability to correctly 'cancel' field contributions when both electric and magnetic fields act on particles. Following the velocity updates, we calculate the energy for each macro particle. Then, their positions are updated according to \eqRef{eq:particle_position}.

With updated positions and velocities, the simulation is prepared for the next and final phase – deposition of particle data back to the mesh. This stage involves computing the current density, which is essential for updating the electric field in the next timestep. Additionally, other mesh variables, such as mass density and bulk momentum, are computed from particle deposition. 


\subsubsection{Current Density Calculation}
\label{sec:charge_density}
Current density calculation is a critical step in ensuring the accuracy of PIC simulations. Often, PIC codes employ a 'charge conserving' method. For example, the PhotonPlasma uses an extension of the charge conserving method developed by \citet{esirkepovExactChargeConservation2001}. This method involves calculating the divergence of the current density by examining how the deposition weights for each macro particle change between timesteps. Following the determination of the divergence, the actual current density is obtained through a prefix sum operation. 

In DISPATCH, such prefix sum operation would require communication across patches to align ghost-values with neighbor patches. The current communication scheme does not allow this, and we therefore opted for a simpler and more direct current density calculation: 

\begin{equation}
    \label{eq:simple_J}
    \mathbf{J} = \sum_s q^s \sum_i w_i \mathbf{v_i} W(x_p - x_c),
\end{equation}
where $W(x_p - x_c)$ represent deposition weight function, $w_i$ the weight of the macro particle, $\mathbf{v_i}$ the velocity of the macro particle, and $q_s$ the charge of the particle species. This method calculates current density directly from the macro particles' properties.

While not exactly charge conserving, down to each macroparticles gradual contribution, the method is nevertheless on average charge conserving, with errors expected mainly on cell scales, where E-field cleaning (Eqs.\ \ref{eq:PhiE}-\ref{eq:EPhi}) can easily take care of making the electric field exactly consistent with actual charge distribution.

\subsection{Particle Sampling}
\subsubsection{Initial Condition Sampling}

\begin{figure}
    \centering
    \includegraphics[width=0.45\textwidth]{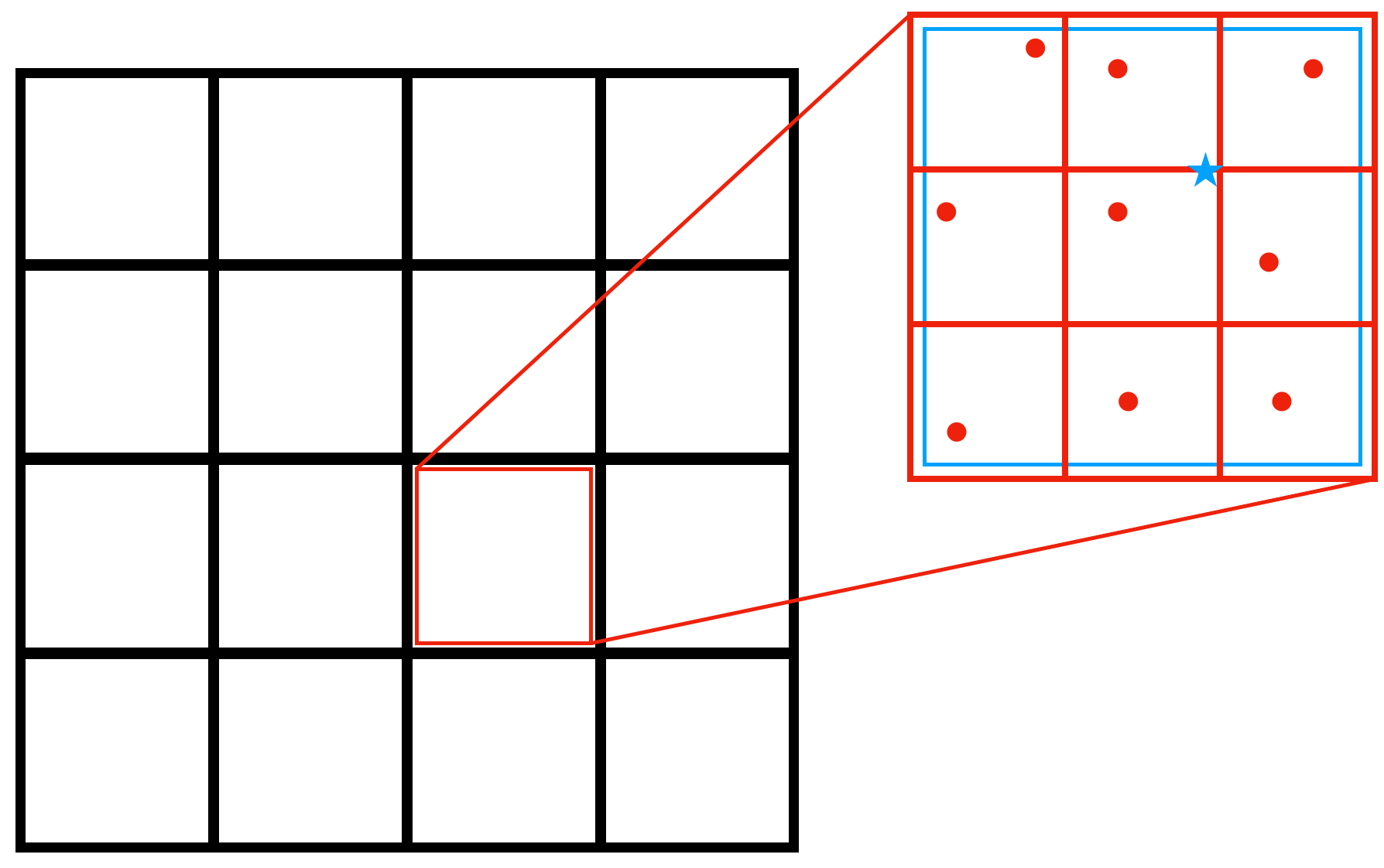}
    \caption{Illustration of particle sampling in a 2D setup with 10 particles per cell. Each cell is divided into \( \lfloor \sqrt{10} \rfloor = 3 \) by 3 sectors, totaling 9 sectors. A particle is placed within each sector (indicated by the red dots). The remaining \( 10-9 = 1\) particle is randomly positioned within the entire cell, as shown by the blue star. }
    \label{fig:position_sampling}
\end{figure}

The initial sampling of particles within our simulation is based on mesh variables such as mass density, current density, bulk velocity, and temperature. These mesh variables, critical for defining initial conditions, are aligned with those used in MHD simulations, to streamline PIC-MHD integration efforts.

To distribute particles evenly at the start, we segment each cell into a predefined number of sectors. The number of sectors, \( n_{\text{sec}} \), varies with the dimensionality of the simulation:
\begin{itemize}
    \item In 3D, \( n_{\text{sec}} = \lfloor n_{\text{pc}}^{1/3} \rfloor \),
    \item In 2D, \( n_{\text{sec}} = \lfloor n_{\text{pc}}^{1/2} \rfloor \),
    \item In 1D, \( n_{\text{sec}} = n_{\text{pc}} \),
\end{itemize}
where \( n_{\text{pc}} \) denotes the number of particles per cell.

For example, in a 2D setup with 10 particles per cell, \( n_{\text{sec}} = 3 \), leading to each cell being divided into \( 3 \times 3 = 9 \) sectors. One particle is placed randomly within each sector, with any additional particles randomly located within the entire cell, as depicted in \figRef{fig:position_sampling}.

After establishing positions, particles' velocities are sampled from a Maxwellian distribution. The initial temperature for this distribution is either provided directly or derived from the internal or total energy, depending on the initial conditions.

Additionally, we incorporate bulk and current velocities into the velocity assignment for particles, similar to the approach described by Daldorf et al. (2014):
\begin{equation}
    \mathbf{v_{elec}} = \mathbf{u_{bulk}} + \frac{m_i}{q_e} \frac{\mathbf{J}}{\rho},
\end{equation}
\begin{equation}
    \mathbf{v_{proton}} = \mathbf{u_{bulk}} + \frac{m_e}{q_i} \frac{\mathbf{J}}{\rho},
\end{equation}
where \(\mathbf{u_{bulk}}\) represents the bulk velocity, \(\mathbf{J}\) is the current density derived from \( \mathbf{\nabla} \times \mathbf{B}\), \(\rho\) is the mass density from grid, and \(m_i, q_i, m_e, q_e\) are the masses and charges of protons and electrons, respectively. 


\subsubsection{Resampling}
\label{sec:resample}

Particles within the inner domain of a patch can freely move between cells. Over time, this freedom can lead to significant discrepancies in the number of particles per cell, with some cells having too many particles and others too few. Similarly, particle distribution discrepancies can occur at the patch level. To maintain a roughly constant number of particles per cell and ensure workload balance across patches, we've implemented a resampling strategy that involves "splitting" particles in underpopulated cells and "merging" particles in overcrowded ones.

Splitting, or up-sampling, is straightforward. The heaviest particle in a cell is divided into two particles, each retaining the original velocity but carrying half the weight. To prevent creating identical particles, their positions are slightly offset. This offset is randomly determined in each dimension from a Maxwell distribution with \(\sigma = 0.1\), with one particle receiving half the offset subtracted and the other half the offset added. This process can then be repeated until the desired number of particles per cell is achieved.

Merging, or down-sampling, presents more complexity, as it can potentially violate conservation laws for mass, energy, and momentum. Drawing on strategies discussed by \cite{muravievStrategiesParticleResampling2021}, we've adopted the 'leveling' approach from their work and introduced a 'close-phase' strategy. For the latter, we evaluate particle velocities within the same cell using the Euclidean distance metric, merging the two particles closest in velocity space into one. This new particle's energy is then compared to the total energy of the original pair. If the energy discrepancy falls within an acceptable tolerance, the merge is finalized; otherwise, the particle is discarded. This process repeats until the target particle count is achieved or all potential merges are explored without success due to excessive energy differences.

The frequency and activation of resampling are left to user discretion, offering flexibility in simulation management.

\subsection{Optimization}
Optimization is crucial for our goal of developing a comprehensive and realistic solar flare model across multiple scales. We've introduced several optimizations to boost the performance of our simulation framework, preparing for future enhancements.

Our simulation adopts a Particle-in-Patch (PIP) approach, similar to DISPATCH's strategy of updating patches instead of individual cells. This method optimizes cache usage and enables effective vectorization of computations. In our PIC solver, each patch hosts several arrays of particle information, facilitating vectorization. Though these arrays are too large for cache, making them less than ideal for traditional computing, their structure is ideal for GPU execution. GPU acceleration remains a key development goal, given its potential to significantly enhance simulation performance.

A fundamental part of our optimization strategy involves sorting particles to improve memory access patterns. When particles spatially close to each other are also adjacent in memory, the efficiency of interpolation and deposition operations increases markedly. We store particle positions in two arrays: an integer array for cell location and a floating array for position within the cell. This dual-array setup simplifies sorting, allowing for integer-based comparisons that improve performance and numerical precision.

The choice of sorting algorithm impacts performance. We have implemented both insertion sort and quick-sort algorithms \citep{CLRS_algorithm_book}. Given the minimal expected changes in particle ordering between timesteps, insertion sort seemed the likely choice. However, initial results indicate better performance with a complete re-sort using quick-sort, even when sorting every timestep. 


Interpolation and deposition are, especially for higher-order schemes, the most demanding processes in our PIC solver, due to inefficient memory access patterns when accessing cells in higher dimensions. Sorting particles addresses this issue to some extent. In our effort to optimize the interpolation routine, we found a more efficient method than interpolating for each particle during the particle pusher step. Instead, interpolating field values for all particles prior to the particle pusher step and storing these values in temporary arrays performed better.

\section{Code Validation}
To validate the PIC solver integrated within the DISPATCH framework, we have devised a comprehensive suite of test cases that progressively increase in complexity. Initially, we conduct a series of unit tests aimed at examining specific equations governing particle motion and field evolution in isolation, i.e., without any feedback between particles and fields. Subsequently, we introduce complexity by incorporating particle-field feedback mechanisms in both 1D and 2D test scenarios. All tests described herein uses CGS units for consistency, and we did not activate resampling to preserve the integrity of conserved variables. 

\subsection{Unit tests}
\subsubsection{Constant Particle Motion}

\begin{figure}
    \centering
    \includegraphics[width=0.45\textwidth]{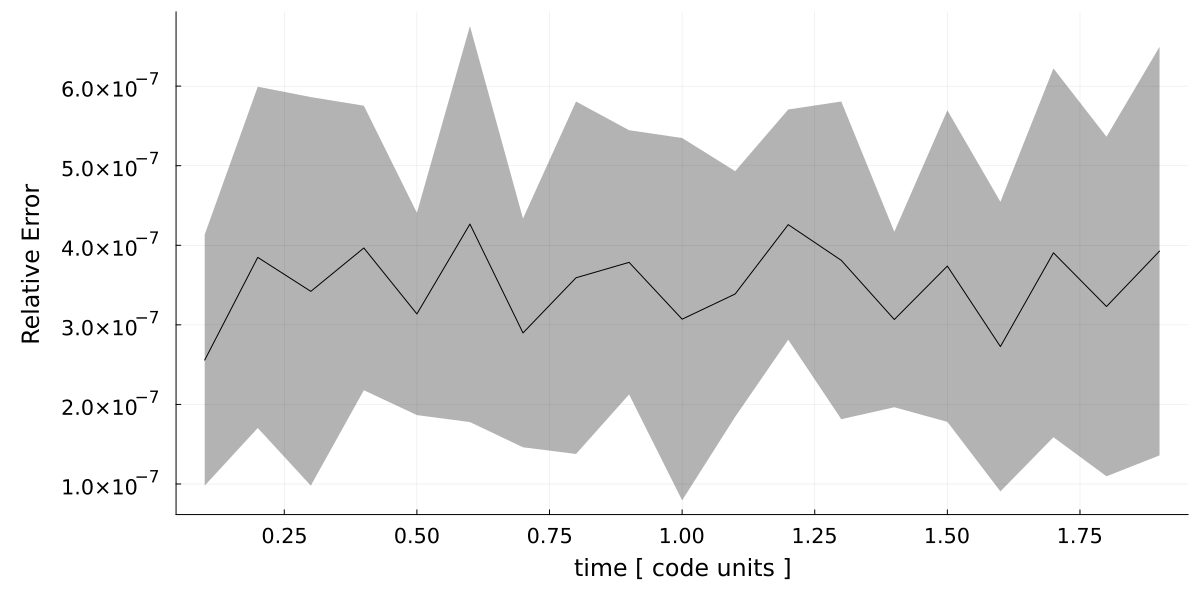}
    \caption{Relative error for the 'single particle motion' test. The solid line shows the average error across the 18 test particles with the shaded area representing the standard deviation. In this test, no fields were present, ensuring that the focus is solely on the accuracy of \eqRef{eq:particle_position} across patch boundaries.}
    \label{fig:unit_test}
\end{figure}

This test, designed to exclusively evaluate \eqRef{eq:particle_position}, initiates with 18 particles, each with distinct velocities across every direction. Both electric and magnetic fields are nullified, preventing any particle-induced field alterations. The experiment is conducted within a 3D domain, comprising a grid of 3x3x3 patches, each containing 8x8x8 cells, to examine the solver's three-dimensional computational integrity. The particle trajectories are tracked to detect any discrepancies as they traverse patch boundaries. \FIGRef{fig:unit_test} illustrates the outcome of this test, affirming that all observed errors remain within the bounds of numerical precision.

\subsubsection{Constant Particle Acceleration}
\begin{figure}
    \centering
    \includegraphics[width=0.45\textwidth]{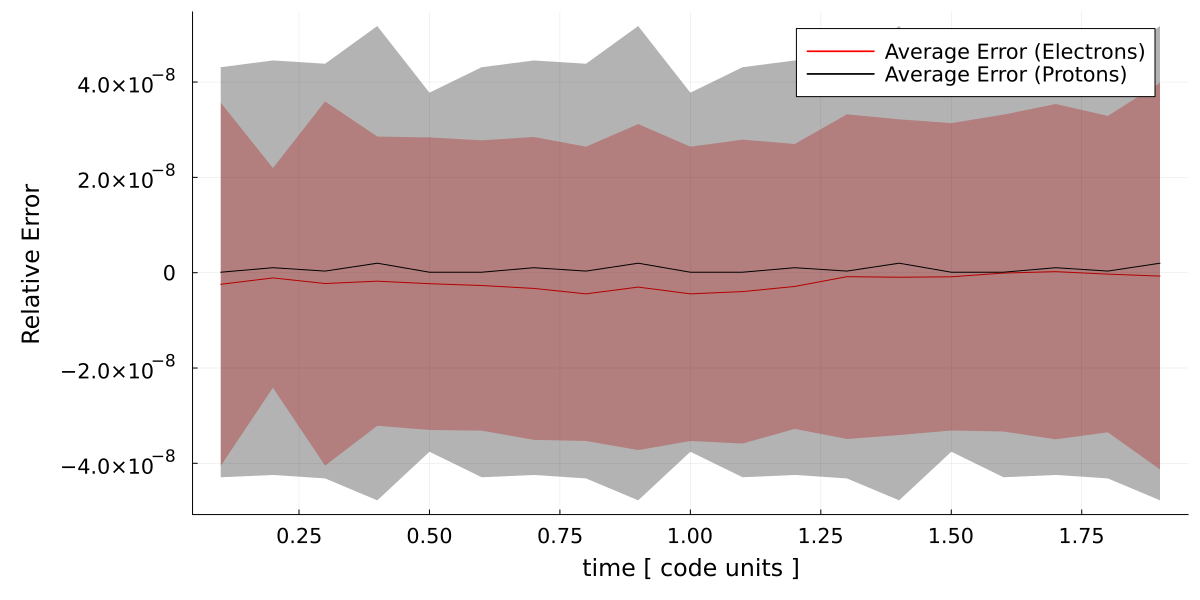}
    \caption{Relative error in velocity constant particle acceleration test. The solid line shows the average errors across the 7 test particles and across 3 seperate runs, with the shaded area representing the standard deviation.}
    \label{fig:const_acc_unit}
\end{figure}

This test specifically aims to validate \eqRef{eq:Lorentz_foce} in the absence of a magnetic field, focusing on the effects of an electric field on particle acceleration. We initialize the experiment with seven particles, each moving along one of the Cartesian axes (X, Y, and Z) in positive and negative direction. The seventh particle has initial zero velocity. To evaluate the solver's accuracy in calculating particle acceleration due to an electric field, we conduct three separate runs. In each run, a constant electric field is applied along one of the axes.

The setup for this test mirrors the configuration used in the previous section, utilizing a 3D grid of 3x3x3 patches, where each patch consists of 8x8x8 cells. In each of the three runs, we analyze the change in proper velocity of the particle moving along the axis aligned with the electric field. The primary objective is to confirm that the solver accurately captures the expected velocity change between snapshots, with a focus on the incorporation of relativistic effects. All observed changes in velocity fall within the realm of numerical precision for both electrons and protons as shown in \figRef{fig:const_acc_unit}.

\subsubsection{E cross B drift}
\begin{figure}
    \centering
    \includegraphics[width=0.45\textwidth]{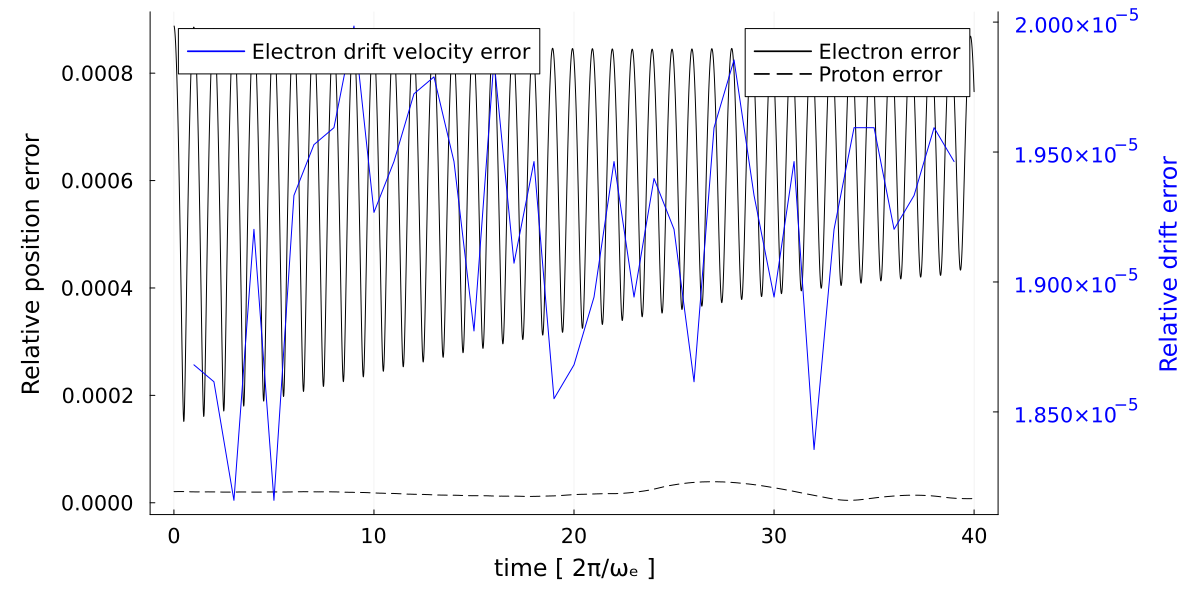}
    \caption{Relative position error for an electron and a proton, alongside drift velocity error for the electron in an E cross B drift test case. Both drift velocity error and positional discrepency remains within numerical precision for over 40,000 timesteps.}
    \label{fig:ExB_drfit}
\end{figure}

The E cross B drift test extends our exploration of \eqRef{eq:Lorentz_foce} to scenarios involving both electric and magnetic fields positioned perpendicularly to each other. This orthogonal arrangement results in a drift velocity expressed as $v_{drift} = c \frac{E_y}{B}$. Specifically, with an electric field oriented in the y-direction ($E_y$) and a magnetic field in the z-direction ($B_z$), the theoretical positions of particles can be determined as follows:
\begin{equation}
    x = x_0 + v_{sign} \cdot r_L \cdot \sin(\omega \cdot t) + \frac{c E_y}{B_z} t,
\end{equation}
\begin{equation}
    y = y_0 + v_{sign} \cdot q_{sign} \cdot r_L \cdot (\cos (\omega \cdot t) - 1),
\end{equation}
where $r_L$ denotes the Larmor radius within the reference frame where $E_y = 0$ — namely, the frame moving at the drift velocity $v_{drift} = \frac{c E_y}{B_z}$. The variable $v_{sign}$ represents the sign of the initial velocity in this specific frame, while $q_{sign}$ indicates the charge sign of the particle, and $\omega$ denotes the gyro frequency.

\FIGRef{fig:ExB_drfit} presents the results from this test. Notably, the figure illustrates that the drift velocity error for electrons stays within the bounds of numerical precision up to $t=40$, equating to around 40,000 timesteps.


\subsubsection{EM-wave in Vacuum}
\begin{figure}
    \centering
    \includegraphics[width=0.45\textwidth]{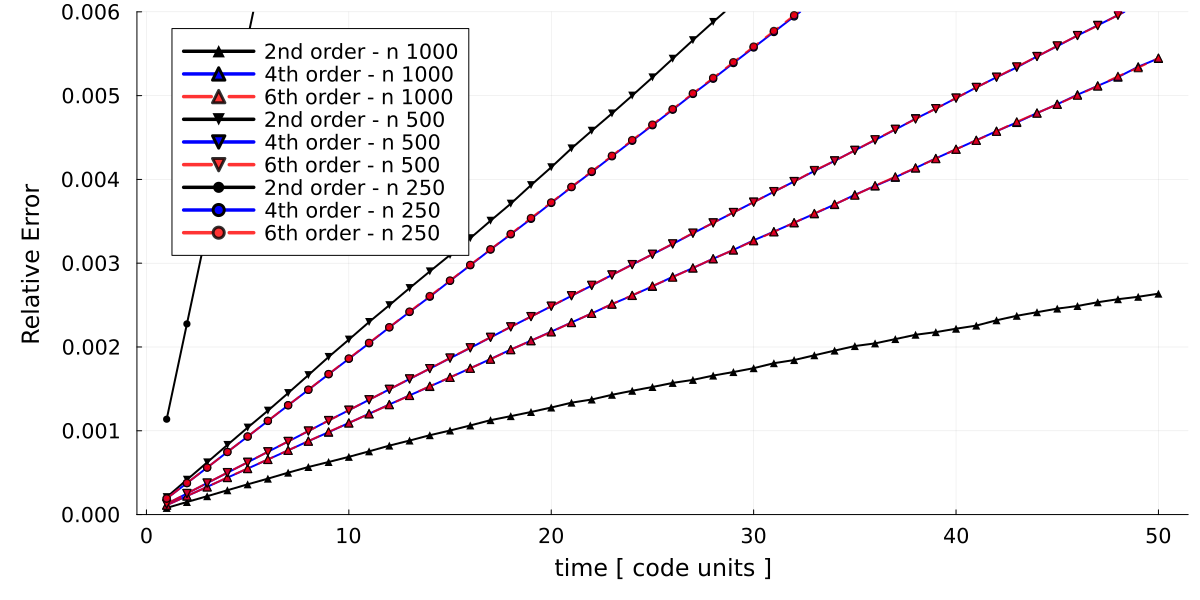}
    \caption{Relative errors in field components, showcasing a sub-linear escalation ascribed to phase shift and amplitude variances. The 4th and 6th order errors overlap, appearing as a single line.}
    \label{fig:EM_wave}
\end{figure}

This test assesses the solver's implementation of \eqRef{eq:Ampere_law} and \eqRef{eq:Faraday_law} in a vacuum scenario with zero current density ($\mathbf{J}=0$). Executed within a 1D setup, the experiment examines the propagation of an electromagnetic wave along a single axis. Separate runs were conducted for each of the three spatial directions for a comprehensive validation. Here, we detail the results for the z-direction. The theoretical model of the wave is described by the following set of equations:
\begin{eqnarray}
    Bx &=& - B_0 \cos(k z - \omega t) \\
    By &=& B_0 \cos(k z - \omega t) \\ 
    Bz &=& 0 \\
    Ex &=& B_0 \cos(k z - \omega t) \\
    Ey &=& B_0 \cos(k z - \omega t) \\
    Ez &=& 0 
\end{eqnarray}
where $k$ represents the wavenumber and $\omega$ the frequency of the electromagnetic wave.

We initialize with $k = \frac{2 \pi \cdot m_0}{L}$, with $L$ indicating the domain's extent and $m_0$ the designated number of wavelengths within this domain. The frequency, $\omega$, is derived using $\omega = k \cdot c$.

The domain's span is set to $1000$ cm, and $m_0 = 2$. Scales are adjusted so that the domain's length corresponds to $1$ in code units, and similarly, time is scaled to set the speed of light, $c$, to $1$ in these units. Various cell counts, including $250$, $500$, and $1000$, were tested to gauge the solver's accuracy across different resolutions. \FIGRef{fig:EM_wave} displays the relative errors encountered in these tests, with derivatives of 2nd, 4th, and 6th orders being examined. Errors predominantly arise from phase shifts and amplitude variations, with phase-shift discrepancies dominating in higher-order cases, while amplitude errors dominate in the 2nd order outcomes.

Error convergence for each order of derivation was critically analyzed. The 2nd order tests showed a 'linear' reduction in errors on a log-log scale, with a slope approximating $-2.2$, aligning closely with the expected $-2.0$ slope indicative of its second-order nature. In contrast, 4th and 6th order tests revealed a diminished slope of $-0.65$, hinting at marginal gains from enhanced resolution. This outcome suggests that the 2nd order time integration of the leapfrog scheme emerges as a significant constraint, limiting the benefits of higher-order spatial resolutions.

To avoid indexing errors in proximity to patch boundaries requires increasing the number of guard zones when using higher-order derivatives, thus incrementally increasing the computational workload. Considering also that a change from 4th to 6th order derivatives does not markedly improve precision, while significantly inflating the computational demands, we conclude that lop-sided increase of only spatial order is not a good idea.

\subsection{Recovery of Plasma frequency}
\begin{figure*}
    \centering
    \includegraphics[width=0.9\textwidth]{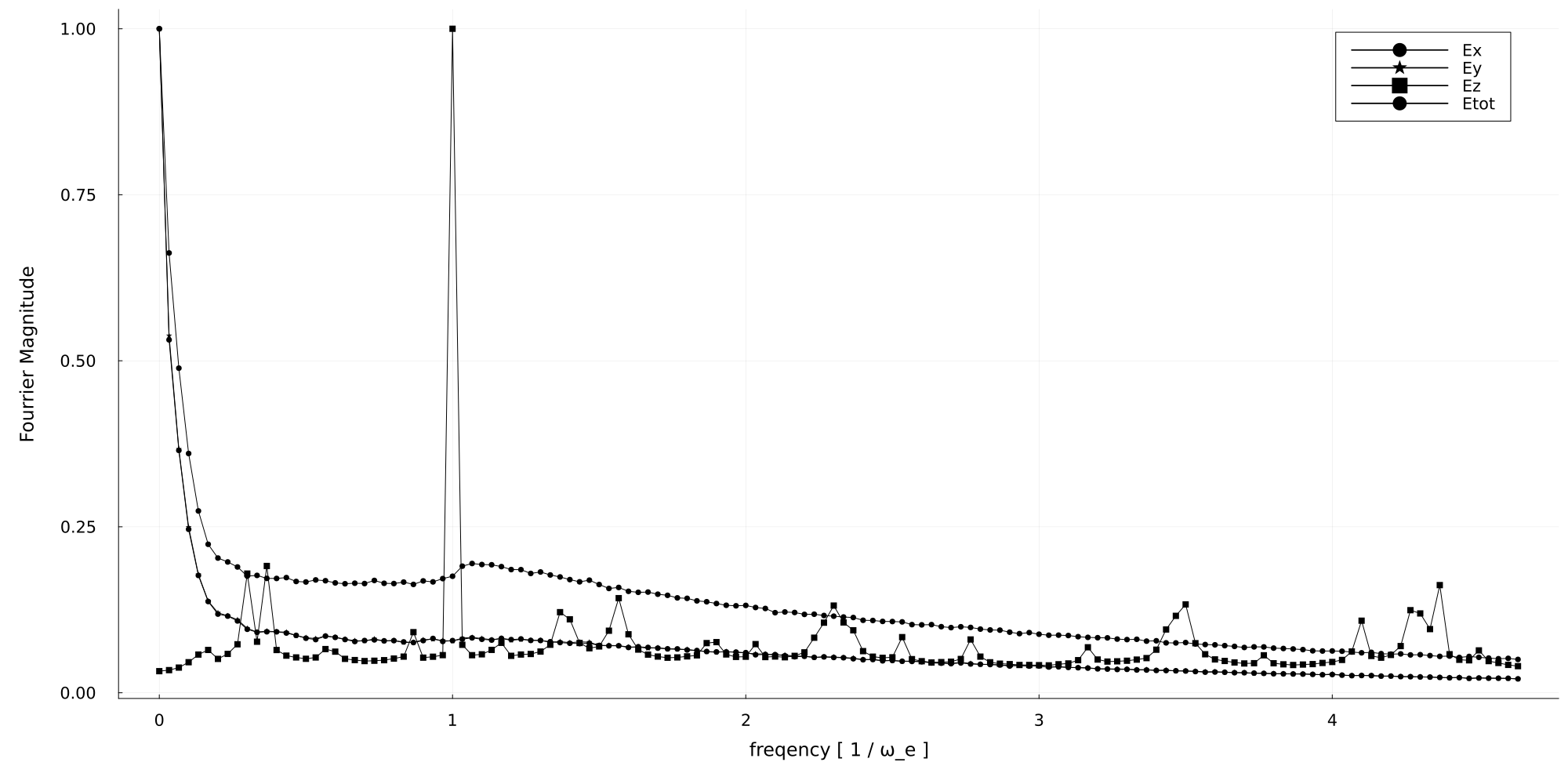}
    \caption{Fourier spectra of electric field components for the plasma frequency experiment, demonstrating clear peaks at a frequency of $1\omega_e$, indicative of accurately capturing the plasma frequency.}
    \label{fig:plasma_freq}
\end{figure*}

The plasma frequency is pivotal in characterizing plasma dynamics. Accurately capturing this frequency in simulations is crucial for validating the scalability and feedback mechanisms within our solver. We explored this through Fourier analysis in a controlled 2D setup, aiming to recover the plasma frequency.

The experimental setup involved initializing electrons and protons at a uniform density of $10^9 \, \text{cm}^{-3}$ within a $5\, \text{cm} \times 5\, \text{cm}$ domain. The domain is segmented into 5x5 patches with 10x10 cells in each for a total of 50x50 grid points. We initialized each species with a thermal velocity corresponding to a temperature of $1\, \text{MK}$. With these parameters the cell width is approximately one Debye length.

We scaled mass density to unity, time such that $\omega_e = 2 \pi$, and did not scale lengths.

As depicted in \figRef{fig:plasma_freq}, the Fourrier analysis reveal distinct peaks corresponding precisely to the plasma frequency. Time has been scaled such that the theoretical plasma period is exactly $1$ 'code Hz'. The peak is most notable in the field component orthogonal to the testing plane. This test demonstrates the solver's adeptness at recovering the correct plasma frequency.

\subsection{Two-Stream Instability}

\begin{figure*}
    \centering
    \includegraphics[width=0.9\textwidth]{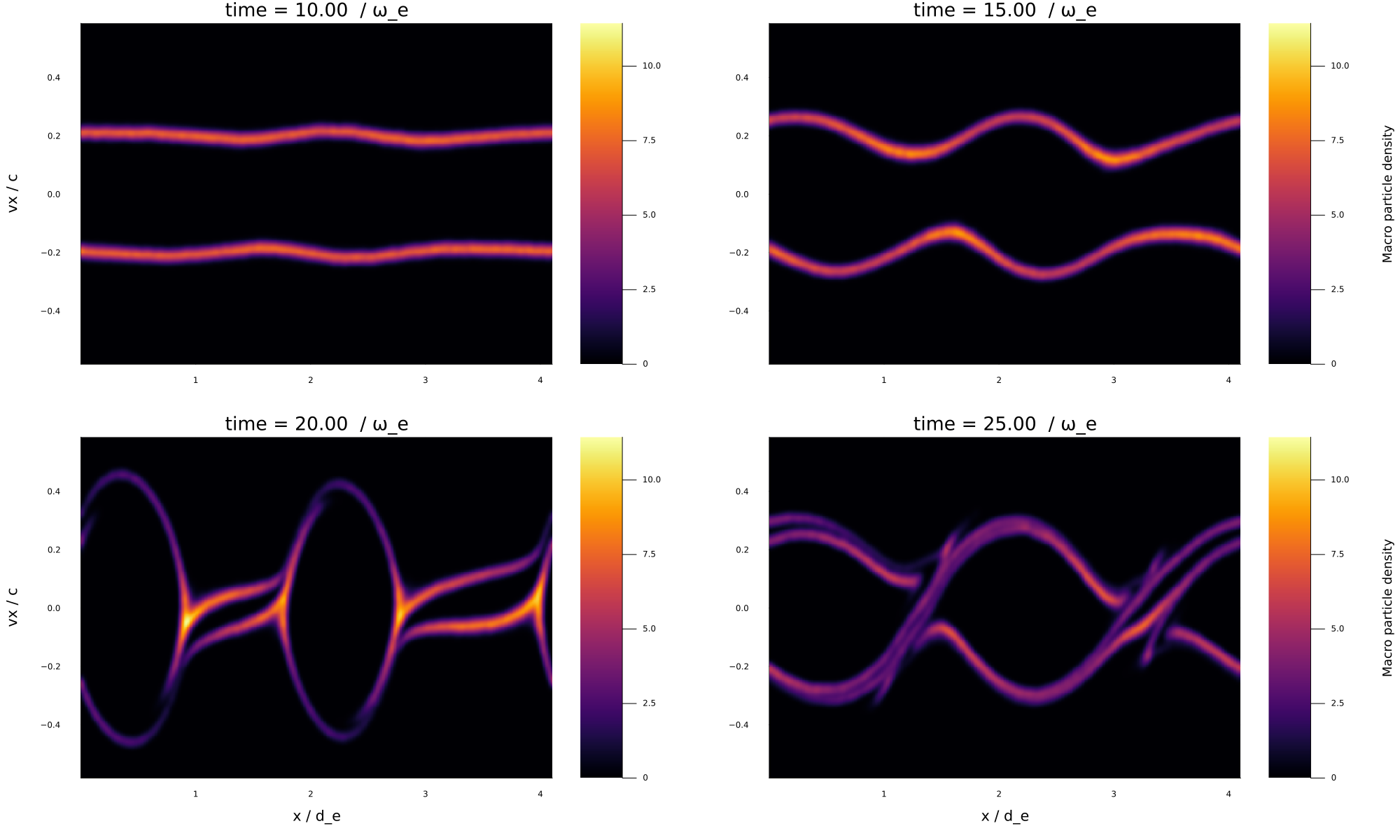}
    \caption{Evolution of the two-stream instability, illustrating macro particle density in phase space at four distinct times, smoothed using a Gaussian filter. The transition from the linear phase to the onset of the nonlinear phase is observable around $t=20 \omega_e$.}
    \label{fig:two_stream_evo}
\end{figure*}

\begin{figure*}
    \centering
    \includegraphics[width=0.95\textwidth]{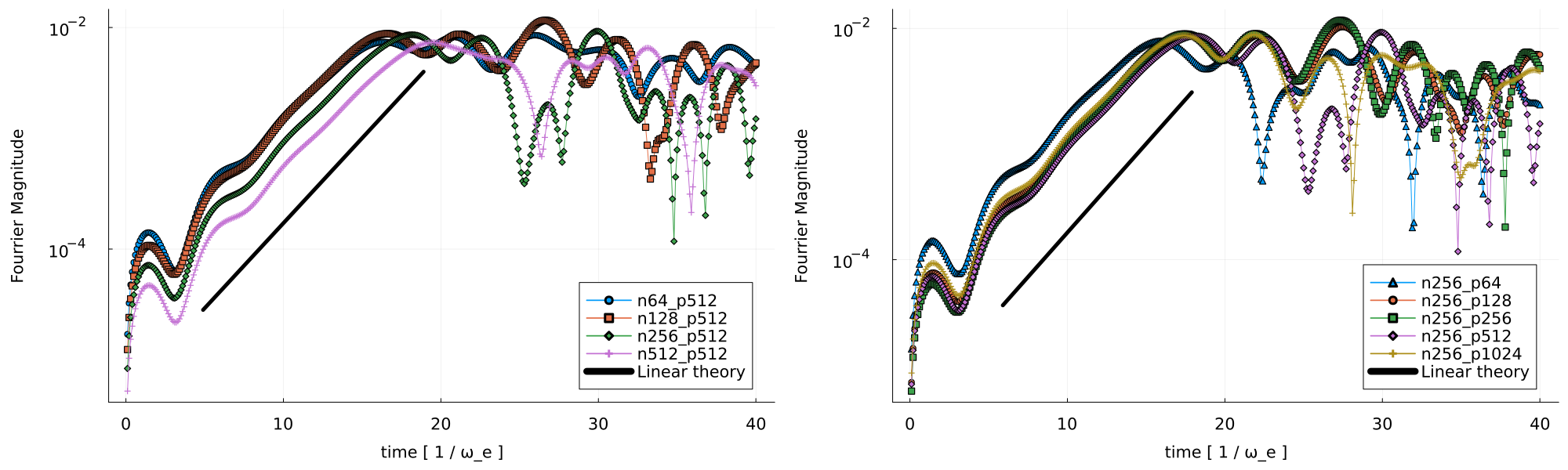}
    \caption{Growth rate of the spectral component $k_{max}$ of the electric field across different simulations, ordered by cell count on the left figure and particles per cell on the right figure.}
    \label{fig:two_stream_growth}
\end{figure*}

Investigating the two-stream instability provides a rigorous assessment of the solver's capability to simulate plasma interactions and instability dynamics. Several different versions of the two stream experiment exists and have been tested  \citep{hazeltineFrameworkPlasmaPhysics1998, cuiGridBasedKineticSimulations2019,jiAsymptoticPreservingEnergyConservingParticleInCell2022a, markidisFluidKineticParticleinCellMethod2014}. Our setup closely resembles \citet{markidisFluidKineticParticleinCellMethod2014}, but is notably different due to different scaling parameters. We initialize our setup with two counter-streaming electron populations against a backdrop of stationary ions. Electrons and protons are initialized with Maxwellian velocities at temperature $T$, and electrons additionally with a drift velocity of $\pm V_0$.

Analyzing the cold case ($T \rightarrow 0$) through the dispersion relation:
\begin{equation}
    0 = 1 - \omega_e^2\frac{1}{2} \left(\frac{1}{(\omega - k V_0)^2} + \frac{1}{(\omega + k V_0)^2} \right) - \frac{\omega_p^2}{\omega},
\end{equation}
yields the maximum growing wave mode at $k_{max} \approx \sqrt{\frac{3}{8}} \frac{\omega_e}{V_0}$, with growth rate $\gamma \approx 0.35355 \omega_e$, when neglecting the contribution from ions. Delving into the warm case revealed the growth rate and mode size changes by less than 1\% for $v_e \leq 0.05V_0$, informing our temperature selection to closely model realistic cold-case plasma conditions.

The computational domain span $4.105 d_e$ where $d_e = \frac{c}{\omega_e}$ represents the electron skin depth. This domain size was chosen to accommodate two wavelengths of the maximum growing mode. We conduct a series of simulations with varying cell counts ($n=[64,128,256,512]$) and particles per cell ($ppc=[64,128,256,512,1024]$). For $n=64$ we used 4 patches with 16 cells in each. For the other cell counts we used 32 cells per patch and $[4,8,16]$ patches, respectively. 


Physical variables were kept unscaled, and we normalized length such that $d_e=1$ in code units and $\omega_e=1$ for time, rendering the speed of light unity in code units as well. We chose a drift velocity of $V_0=0.2c$, which sets the temperature $T\approx 592,990$ K for $v_e = 0.05\ V_0$. Additionally, we used a number density of $10^{10} cm^{-3}$ and scaled mass, such that the mass density is 1 in code units. 

The simulations proceeded without field cleaners or particle resampling until $t=40$. \FIGRef{fig:two_stream_evo} illustrates the evolution of the linear phase and the initial emergence of the nonlinear phase, showing the formation of dual structures in phase space as anticipated from the experiment's design.

\FIGRef{fig:two_stream_growth} shows the electric field's growth rate for the dominant wavemode, illustrating consistency with the anticipated growth rates derived from linear theory. This alignment is consistent across various experimental configurations, underscoring the solver's robustness in accurately modeling the intricate processes underlying two-stream instabilities under a spectrum of simulation conditions.

\subsection{Current Sheet}
\label{sec:current_sheet}

\begin{figure}
    \centering
    \includegraphics[width=0.45\textwidth]{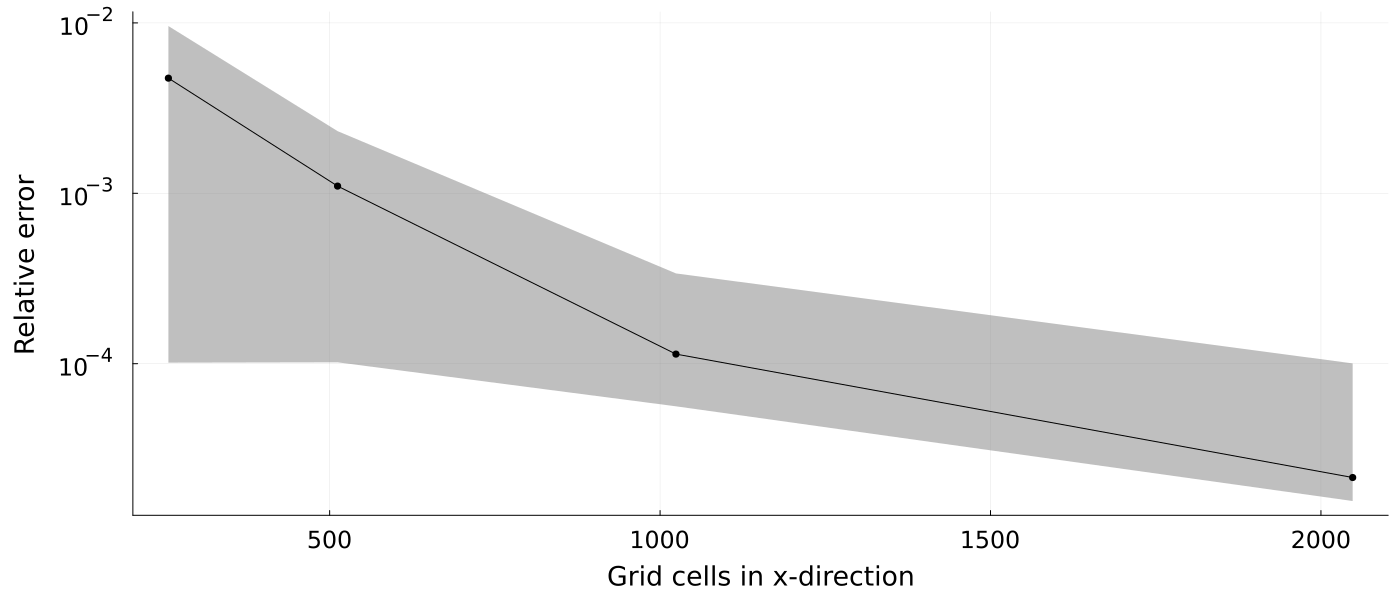}
    \caption{Relative total energy error for c10m400 scaling current sheet runs with 32 particles per cell at simulation's end, averaged over the last ten snapshots. The shaded area shows the standard deviation.}
    \label{fig:energy_error}
\end{figure}

\begin{figure}
    \centering
    \includegraphics[width=0.45\textwidth]{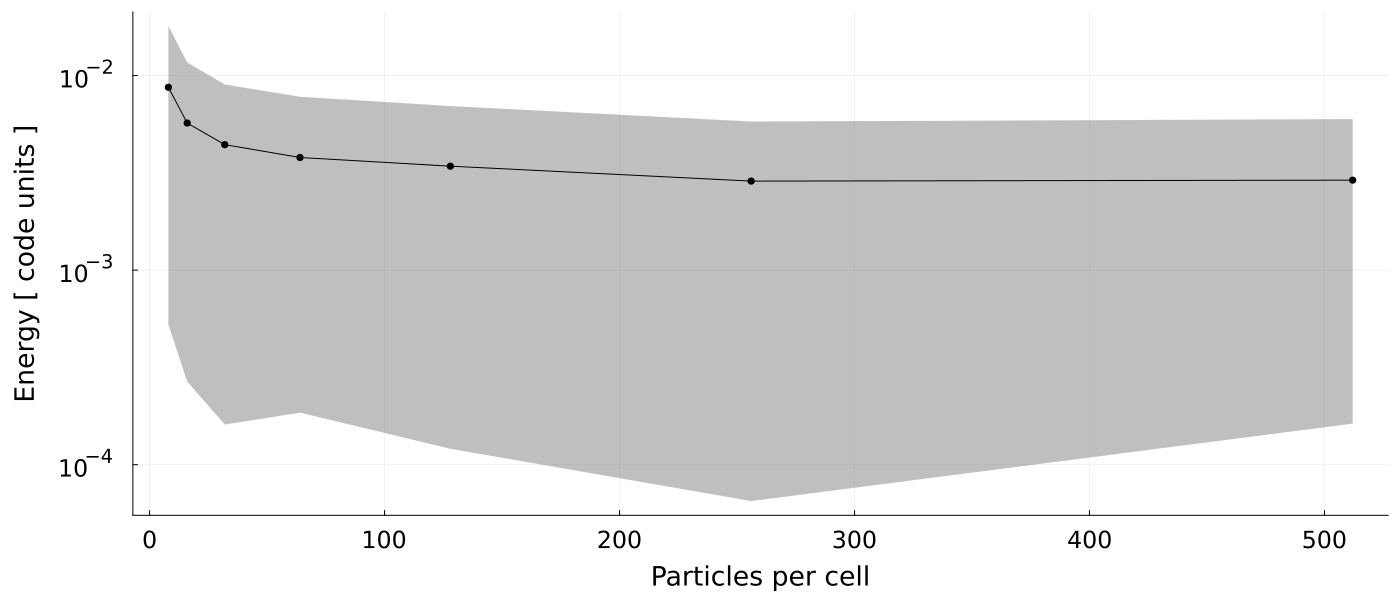}
    \caption{Final electric energy as a function of particle count per cell in c10m400 scaling current sheet scenario at n=256x128 resolution. Values are averages over the last ten snapshots. The shaded area shows the standard deviation.}
    \label{fig:electric_energy}
\end{figure}

The final test is the behavior of the physical effects under different parameter scaling scenarios. We chose a current sheet setup as such a scenario is inherently unstable and should quickly highlight any differences in evolution. Current sheet behavior is an open research topic with many authors investigating current sheet in space and solar contexts, wiht varying degree of realism (e.g.\ \cite{popescubraileanuTwofluidReconnectionJets2023}, \cite{liuFirstPrinciplesTheoryRate2022a}, \cite{jiMagneticReconnectionEra2022}, \cite{daldorffImpact3DStructure2022}). For our initial validation, we have chosen a simpler setup, inspired by \cite{hawleyMOCCTNumericalTechnique1995}. The magnetic field configuration is defined by two hyperbolic tangent functions:

\begin{equation}
B_x(y) = 
\begin{cases} 
- B_0 \text{tanh}\left( \frac{y + 3.6}{0.5 \cdot L} \right) & \text{if } y < 0, \\
B_0 \text{tanh}\left( \frac{y - 3.6}{0.5 \cdot L} \right) & \text{if } y \geq 0,
\end{cases}
\end{equation}
where $B_0 = 50$ G represents the background magnetic field strength and $L$ the initial width of the current sheets. We set $L=r_p$ where $r_p$ is the proton gyroradius. The simulation initializes electrons and protons with Maxwellian distributions at $T=1$ MK and uniform density $n=10^9 \text{cm}^{-3}$ within a domain sized $[25.6,12.8]r_p$. Periodic boundary conditions are applied along both the x and y axes. An initial velocity perturbation:
\begin{equation}
    v_y = v_0 \cdot \text{sin} \left( \frac{2 \pi}{L_x} x \right),
\end{equation}
with $v_0 = 0.001\ c$, stimulates the onset of reconnection.

Experimentation varied over cell counts ([128x64, 256x128, 512x256, 1024x512, 2048x1024]) and particles per cell (\texttt{ppc} = [32, 64, 128, 256]). For \texttt{n} = [128x64, 256x128] we used 16x16 cells per patch and [8x4,16x8] patches. For \texttt{n} = [512x256, 1024x512, 2048x1024] we used 32x32 cells per patch and [16x8, 32x16, 64x32] patches.

Four distinct 'fudging' strategies were applied to probe their influence on simulation behavior:


\begin{itemize}
    \item Pure Runs: Physical variables remained unaltered, providing a control scenario for comparison.
    \item m400 Runs: Electron mass was scaled to set the mass ratio $m_p / m_e = 400$.
    \item c10 Runs: Speed of light was reduced tenfold, achieved by augmenting $\epsilon_0$ by 100.
    \item c10m400: Combined the adjustments from m400 and c10 runs for cumulative effect analysis.
\end{itemize}

Modifying the elementary charge primarily reduces the gap between microscopic and macroscopic scales \citep{baumannKINETICMODELINGPARTICLE2013a}. In a pure PIC simulation, changing the elementary charge only affects the length and time scales. When scaled to code units, there is no difference between runs with and without elementary charge scaling. This is unlike scaling the electron mass or the speed of light, which explicitly changes the ratios between parameters of interest, even in code units. Therefore, we did not make any adjustments to the elementary charge.

We kept scaling consistent across scenarios for analytical clarity. We scaled length such that $r_p = 1$, time such that plasma frequency $\omega_e = 2 \pi$, and initial mass density so $\rho_0 = 1$ in code units for the non-fudged run.

\begin{figure*}
    \centering
    \includegraphics[width=0.95\textwidth]{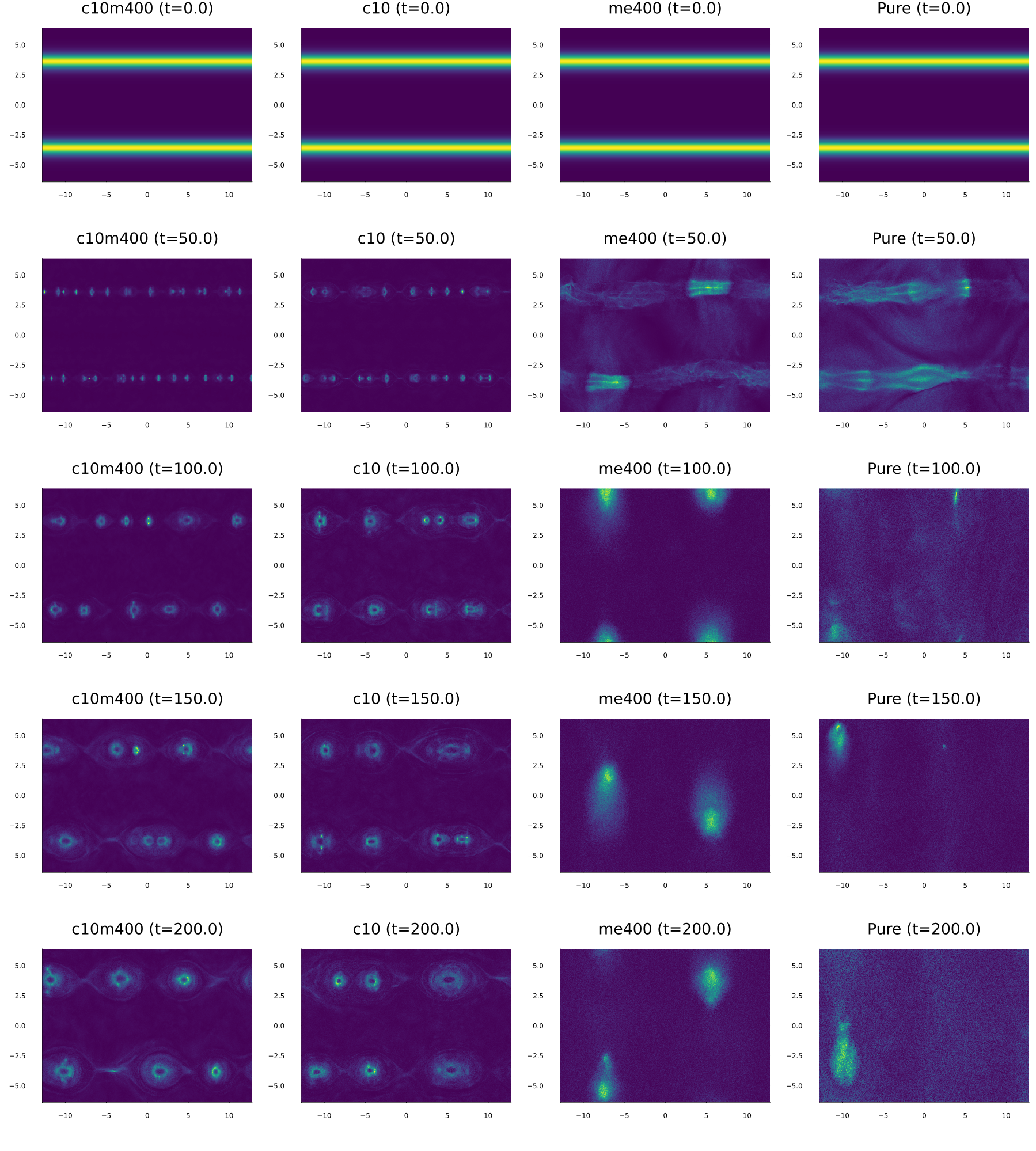}
    \caption{Comparative evolution of the total current density for the four fudging scenarios at selected times, using a resolution of $1024x512$ cells with 32 particles per cell. Differences in plasmoid formation highlight the effects of scaling adjustments.}
    \label{fig:J_comparison}
\end{figure*}


The current density evolution depicted in \figRef{fig:J_comparison} illustrates the effects of electron mass and light speed adjustments on plasmoid dynamics. Particularly, the c10 and c10m400 runs showcase a prevalence of smaller-scale plasmoids, contrasting with the more homogeneous current dissipation in pure and m400 scenarios.

Figure \ref{fig:energy_error} illustrates convergence towards zero relative total energy error with increasing resolution, in alignment with theoretical expectations for a second order solver. A slope of $-2.2$ was found, similar to our findings from the EM-wave test.

The influence of the particle count per cell on electric field energy stabilization is investigated in \figRef{fig:electric_energy}, affirming the solver's stability under cell sizes below 10 Debye Lengths. Theoretically, $\Delta x < 3.4 \lambda_D$ must be satisfied to avoid self-induced instabilities \citep{tskhakayaParticleCellMethod2007}. The theoretical analysis assumes that the number of particles inside the Debye radius is large. In our experiment, we show that the solver is stable for as little as 2 particles per Debye area for 2D simulations.

This comprehensive suite of tests validates the solver's robustness in modeling the intricate dynamics of current sheet formation and magnetic reconnection, setting a solid foundation for expanding investigations into 3D simulations.

\section{Discussions and conclusion}
\subsection{Speed of light and related issues}
The integration of the explicit PIC solver within the DISPATCH framework marks our first step forward towards simulating the complex plasma dynamics of solar flares. This integration also highlights certain limitations inherent to the nature of explicit solvers, and how this plays out relative to the DISPATCH code framework. 

A primary challenge lies in the limitations on the timestep size by the speed of light; a consequence of the need to accurately model electromagnetic field propagation. This constraint, while potentially limiting in other, less extreme contexts, aligns well with the solver's intended application to scenarios involving relativistic particle acceleration, such as observed in solar flares.  One can hardly claim to realistically model such scenarios without embracing and accepting that one must deal with the full range of velocities.

Moreover, relativistic conditions offer other advantages, such as effectively reducing the impact of the difference of mass between electrons and ions, with both species moving at essentially the same velocities, despite large difference in mass.

The explicit solver's reliance on uniform timesteps might seem to diverge from DISPATCH's otherwise innovative approach of employing local time stepping. To some extent, this is an unavoidable consequence of the fundamental impact of the speed of light in relativistic dynamics---when information speed is actually essentially constant there is no advantage to gain from variations in speed, but the framework still allows for coping with large differences in time steps due to differences in spatial resolution, coming from fixed or adaptive mesh refinement.

\subsection{Exposing the consequences of `fudging'}
As previously mentioned, the 'fudging' of physical constants is a key component in many PIC codes. However, the descriptions of these adjustments are often unclear. With our 'transparent scaling,' we address this issue by clearly outlining the modifications and their implications.

Some examples of the impact of fudging are examined in Section \ref{sec:current_sheet}. We demonstrate how varying the speed of light and the mass of the electron alters the fundamental plasma scales. Specifically, reducing the speed of light by a factor of 10 increases the allowable timestep by the same factor. However, simultaneously, the relevant scales are reduced by approximately a factor of 10, and thus the benefits offered by this fudging might be partially or fully offset. As illustrated by \figRef{fig:J_comparison}, the results with and without fudging may differ dramatically, and thus the methodology requires very careful vetting of assumptions and consequences, as made possible by the clear distinction made here between fudging on the one hand, and scaling to code units and choice of unit system on the other hand.

\subsection{Conclusion}
We have successfully introduced an explicit PIC solver into the DISPATCH framework, adapting and building upon the PhotonPlasma code's foundations. 

Our comprehensive validation process, encompassing both straightforward unit tests and more intricate simulations such as the double current sheet test, has rigorously confirmed the PIC solver's robustness and precision. These efforts underscore its capacity to simulate the nuanced interplay of particle dynamics and electromagnetic fields, offering a robust platform for cutting-edge research in solar and plasma physics.

The integration into the DISPATCH code framework not only broadens the solver's applicability across diverse simulation scenarios but also lays the groundwork for a hybrid PIC-MHD implementation. Such an implementation will greatly extend our ability to model solar flares as well as other complex astrophysical phenomena with unprecedented realism, by using the cheaper MHD representation where possible, switching to the much more expensive PIC representation only where needed. 

\begin{acknowledgements}
We would like to thank Troels Haugbølle for his invaluable discussions and insights regarding the PhotonPlasma code and plasma physics. We also wish to acknowledge Jan Truelsen for his assistance and contributions to the theoretical foundations of this work, particularly insight into the warm case analysis of the two-stream instability test. Additionally, we thank Andrius Popovas and Mikolaj Szydlarski for their guidance and support throughout the code development process.

This research was supported by the Research Council of Norway through its Centres of Excellence scheme, project number 262622.
\end{acknowledgements}

\bibliographystyle{plainnat}  
\bibliography{refs}  

\end{document}